\documentclass[sn-nature]{sn-jnl}
\usepackage{graphicx}%
\usepackage{multirow}%
\usepackage{amsmath,amssymb,amsfonts}%
\usepackage{amsthm}%
\usepackage{mathrsfs}%
\usepackage[title]{appendix}%
\usepackage{xcolor}%
\usepackage{textcomp}%
\usepackage{manyfoot}%
\usepackage{booktabs}%
\usepackage{algorithm}%
\usepackage{algorithmicx}%
\usepackage{algpseudocode}%
\usepackage{listings}%
\usepackage{longtable}
\usepackage{anyfontsize}

\newcommand{\Bl}{\langle B_{\ell}\rangle}
\newcommand{\Bp}{\langle B_{\mathrm{p}}\rangle}
\newcommand{\BcL}{\langle B_{\mathrm{cL}}\rangle}
\newcommand{\BcM}{\langle B_{\mathrm{cM}}\rangle}

\begin{document}

\title[Article Title]{Measuring the magnetic fields in the chromospheres of low-mass stars}

\author*[1,2]{\fnm{Tianqi} \sur{Cang}}\email{tianqi\_cang@bnu.edu.cn}
\author*[3]{\fnm{Pascal} \sur{Petit}}\email{ppetit@irap.omp.eu}
\author[3]{\fnm{Jean-Fran\c{c}ois} \sur{Donati}}
\author*[4,5]{\fnm{Hui} \sur{Tian}}\email{huitian@pku.edu.cn}
\author*[1,6]{\fnm{Jianning} \sur{Fu}}\email{jnfu@bnu.edu.cn}
\author[6]{\fnm{Hao} \sur{Li}}
\author[3,7]{\fnm{Stefano} \sur{Bellotti}}
\author[1]{\fnm{Xueying} \sur{Hu}}
\author[1]{\fnm{Xiaoyu} \sur{Ma}}
\author[3]{\fnm{Arturo} \sur{Lopez Ariste}}
\author[1]{\fnm{Keyu} \sur{Xing}}
\author[8]{\fnm{Julien} \sur{Morin}}
\author[9]{\fnm{Hongpeng} \sur{Lu}}
\author[1,5]{\fnm{Weikai} \sur{Zong}}

\affil[1]{\orgdiv{School of Physics and Astronomy}, \orgname{Beijing Normal University}, \orgaddress{\street{No.19 Xinjiekouwai Street}, \city{Beijing}, \postcode{100875}, \country{People's Republic of China}}}

\affil[2]{\orgdiv{Department of Scientific Research}, \orgname{Beijing Planetarium}, \orgaddress{\street{No.138 Xizhimenwai Street}, \city{Beijing}, \postcode{100044}, \country{People's Republic of China}}}

\affil[3]{\orgdiv{Institut de Recherche en Astrophysique et Planétologie}, \orgname{Université de Toulouse,CNRS, IRAP/UMR 5277}, \orgaddress{\street{14 avenue Edouard Belin}, \city{Toulouse}, \postcode{31400}, \country{France}}}

\affil[4]{\orgdiv{School of Earth and Space Sciences}, \orgname{Peking University}, \orgaddress{\street{5 Yiheyuan Road}, \city{Beijing}, \postcode{100871}, \country{People's Republic of China}}}

\affil[5]{\orgdiv{State Key Laboratory of Solar Activity and Space Weather, National Space Science Center}, \orgname{Chinese Academy of Sciences}, \orgaddress{\street{NO.1 Nanertiao, Zhongguancun}, \city{Beijing}, \postcode{100190}, \country{People's Republic of China}}}

\affil[6]{\orgdiv{Institute for Frontiers in Astronomy and Astrophysics}, \orgname{Beijing Normal University}, \orgaddress{\street{Shahe, Manjing road}, \city{Beijing}, \postcode{102206}, \country{People's Republic of China}}}

\affil[7]{\orgdiv{Leiden Observatory}, \orgname{Leiden University}, \orgaddress{\street{PO Box 9513}, \city{Leiden}, \postcode{2300 RA}, \country{The Netherlands}}}

\affil[8]{\orgdiv{Laboratoire Univers et Particules de Montpellier}, \orgname{Universit\'e de Montpellier, CNRS}, \orgaddress{\street{Campus Triolet Place Eugène Bataillon – CC 72}, \city{Montpellier}, \postcode{34095}, \country{France}}}

\affil[9]{\orgdiv{State Key Laboratory of Public Big Data and Guizhou Radio Astronomical Observatory}, \orgname{Guizhou University}, \orgaddress{\street{2708 Huaxi Blvd South Section}, \city{Guiyang}, \postcode{550025}, \country{People's Republic of China}}}

\maketitle

\clearpage
\section*{Abstract}
{\bf
Magnetic fields in the upper atmospheres of solar-like stars are believed to provide an enormous amount of energy to power the hot coronae and drive large-scale eruptions that could impact the habitability of planetary systems around these stars. However, these magnetic fields have never been routinely measured on stars beyond the solar system. Through decade-long spectropolarimetric observations, we have now achieved the measurements of magnetic fields in the lower and middle chromospheres of three M-dwarfs.
Our results indicate that the line-of-sight component of the chromospheric magnetic fields can reach up to hundreds of Gauss, whose sign frequently opposes that of the photospheric field. The measurements highlight the magnetic field complexity and the variation with height close to the surface of these M-dwarfs. They provide critical constraints on the energy budget responsible for heating and eruptions of stellar upper atmospheres, and enable assessments of how stellar magnetic activity may affect exoplanet environments.
}
\section*{}
As a layer of the stellar upper atmosphere, the chromosphere lies above the photosphere and beneath the corona\cite{linsky2017stellar}. It serves as the locus of thermal inversion, where the temperature begins to rise after decreasing from the stellar core to the photosphere. This characteristic demarcates it as more than a transitional layer, transforming it into a site with various types of activity \cite{2007Sci...318.1574D, 2014Sci...346A.315T, carlsson2019new, 2024Sci...386...76Y}, which significantly influences the space weather conditions surrounding orbiting planets\cite{2017NatAs...1E.129L, 2021NatPh..17..807L}. The magnetic field predominantly governs such activity, storing and releasing energy, thereby significantly affecting the heating processes and efficiently driving large-scale eruptions in the chromosphere and corona \cite{2014Sci...346E.315H}. 
Consequently, measuring the magnetic field of the upper atmosphere, including the chromosphere, is essential to understanding the physical mechanisms responsible for stellar coronal heating and eruptions\cite{2023MNRAS.526.1646D,2017SSRv..210...37L}.

However, magnetic field measurements in the stellar upper atmosphere are very limited because of the weak signal, even for the Sun\cite{2003Natur.425..692S,2020Sci...367..278F,yang2020global, 2020NatAs...4.1140C, 2016NatPh..12..179J,2020ScChE..63.2357Y}. Although there have been attempts at magnetic field measurements in the chromospheres of accreting pre-main-sequence stars\cite{2007MNRAS.380.1297D, 2011MNRAS.412.2454D}, these signals are intertwined with the accretion process, which has a different energy-transfer mechanism compared to a solar-like chromosphere. In the absence of measurements, our knowledge about magnetic fields in the upper atmospheres of distant main-sequence stars other than the Sun relies on extrapolations \cite{jardine2019slingshot, vidotto2021evolution} of the measured photospheric magnetic fields\cite{donati2009magnetic, reiners2012observations}. However, comparisons between the measured and extrapolated magnetic fields in the solar upper atmosphere often reveal distinct discrepancies, suggesting that the assumptions of extrapolations are invalid at many locations on stars\cite{2005A&A...433..701W, carlsson2019new, yang2020global}. Furthermore, the majority of magnetic energy in stellar atmospheres is stored in small-scale fields\cite{1998Natur.394..152S,2020A&A...635A.142K}, which are crucial for understanding stellar activity mechanisms. Unlike the photosphere, where magnetic structures are more concentrated, chromospheric fields typically exhibit more diffuse patterns\cite{2014A&ARv..22...78W}. Solar observations demonstrate that this structural complexity can lead to varying measurement outcomes across the chromosphere\cite{ishikawa2021mapping}, with magnetic field vectors sometimes appearing in opposite directions between chromospheric and photospheric layers in the same region\cite{2019A&A...621A...1R}.

M-dwarfs are low-mass stars with surface temperatures lower than those of the Sun and extended convective envelopes. They are the most common type of star in the solar neighborhood, making up over 60\% of all stars, and they host the majority of known exoplanets\cite{bonfils2013harps}. With a mass less than $0.6 M_\odot$, M-dwarfs have a lifetime longer than that of the Sun, and the habitable zones of their orbiting exoplanets are closer, often less than $0.3$ astronomical unit (AU) away from the stars\cite{chabrier2000theory}. The probability of detecting habitable planets is higher around M-dwarfs compared to earlier-type stars. However, the habitability is more likely to be influenced by the magnetic activity originating primarily from the host stars\cite{vidotto2015environment,linsky2019host,2021LRSP...18....3V}. Many M-dwarfs exhibit greatly elevated levels of magnetic activity compared to our Sun, exhibiting frequent superflares with total energies of $10^{33-38}$ erg\cite{2014ApJ...792...67C,2022ApJ...933...92C}. While the surface magnetic fields of M-dwarfs have been measured over the past decades\cite{donati2009magnetic, reiners2012observations,2021A&ARv..29....1K}, the energy driving this magnetic activity primarily originates from the magnetic fields in the upper atmospheres, which have not yet been measured\cite{wright2011stellar,reiners2022magnetism}. In addition, many active M-dwarfs are fully convective stars, but it is still debated whether the lack of a tachocline will result in a dynamo different from the solar one. Understanding the key physical processes in these dynamos requires constraints from observations of small-scale magnetic field structures in stellar atmospheres, which are still lacking\cite{2021A&ARv..29....1K}.

Spectral lines originating at different atmospheric heights provide diagnostic windows into distinct stellar layers. While most absorption lines in stellar spectra form within the photosphere, the unique physical conditions in the chromosphere selectively enhance specific spectral features, particularly the Balmer lines and Calcium II lines. These enhanced emission features serve as valuable probes of chromospheric conditions and magnetic structures\cite{linsky2017stellar}. We analyzed a decade-long compilation of polarized spectra obtained with two high-resolution spectropolarimeters, ESPaDOnS and NARVAL (Methods). The data include the unpolarized normalized intensity (Stokes $I$, represented by normalization to the continuum $I/I_c$) and circular polarization (Stokes $V$, represented by $V/I_c$).  This study focuses on three active mid to late M-dwarfs \cite{2024ApJ...960...62E}: AD Leo, YZ CMi, and EV Lac, which have masses of 0.31-0.42 solar mass and rotation periods of 2.2-4.4 days (see Supplementary Information, SI hereafter, Table S1). 

We simultaneously measured the mean longitudinal magnetic field (detailed in Method) of these stars in the photosphere $\Bp$ from mean photospheric lines, the lower chromosphere $\BcL$ from the H$\alpha$ line, and the middle chromosphere $\BcM$ from the cores of the Calcium II infrared triplet (Ca IRT). Figure 1 shows examples of Stokes $I$ and $V$ profiles for each star. The anti-symmetric features in the Stokes $V$ profiles are typical signatures of the Zeeman effect \cite{2004......deglinnocenti}. Figure 2 presents all measurements for the three targets, along with the corresponding rotational phases of the observations\cite{morin2008large,morin2008stable}. The mean longitudinal magnetic fields in the chromospheres of the three targets can reach several hundred Gauss, comparable to the photospheric fields.

\section*{Results}
\paragraph*{Large-scale magnetic fields}
The variation of the measured magnetic fields at different rotational phases can be used to characterize large-scale magnetic field structures\cite{donati1997spectropolarimetric,morin2008large}. We typically obtained one spectrum per observation night for each star, resulting in low sampling coverage within one rotation cycle. Given that the large-scale photospheric magnetic field structure can vary yearly but generally remains stable within an observing epoch (the few-month window each year during which the targets are visible)\cite{2023A&A...676A..56B,2024A&A...686A..66B}, we grouped the data by epochs for further analysis.  
The time duration of one epoch was at least one month, which is much longer than the stellar rotational period. Therefore, our measurements of the magnetic fields were taken over multiple rotation cycles within each epoch. For each target, we focus on measurements from a marked epoch with full rotational phase coverage: AD Leo observed in 2008, YZ CMi observed in late December 2007 and early 2008, and EV Lac in 2007. We neglect the effect of differential rotation, since it is not detected for the photospheric field in the three targets\cite{morin2008large}. In these epochs, we see that observations at different times form relatively stable patterns in a phase-folded diagram as shown in Figure 2 (SI 1B \& Figure S3), which can be fitted by a two-harmonic Fourier series. This stability suggests that the large-scale chromospheric fields co-rotating with the stars were nearly stable within an epoch, with a timescale of at least one month.

\paragraph*{Layer-dependent magnetic fields} The different patterns of layers in the same epoch indicate that the large-scale magnetic fields vary with height. As shown in Figure 2, while the sign of $\BcL$ is generally consistent with $\Bp$, frequent opposite signs in $\BcM$ highlight the different magnetic topologies with height. Among the three stars, AD Leo has the highest percentage of observations where $\BcM$ and $\Bp$ have opposite signs. Example spectra with the opposite sign between photosphere and middle chromosphere can be found in Figure 1 for AD Leo and EV Lac, showing the same orientation of the anti-symmetric structure for $V$ profile and opposite orientation of $I$ profile. Disk-integrated spectral profiles average the magnetic field contributions over the entire visible surface of the star, potentially masking localized mixed-polarity magnetic features but providing a comprehensive view of the star’s overall magnetic structures. Considering that the chromospheric layer is close to the surface of a star (0.3\% R$_\odot$ for Sun), the variation in polarity and strength at different heights can be attributed to the variation of corresponding large-scale magnetic field structures with height, resulting in different contributions of positive- or negative-polarity fields to the observed signals. The sign change can be reproduced by extrapolation of the photospheric field (Figure S4 \& S5), but detailed variation between the patterns should be attributed to more complex effects close to the surface, like small-scale fields or more diffuse large-scale magnetic loops than predicted. This scenario is different from that on the Sun, where the polarity and topology of the magnetic field from the photosphere to the chromosphere are generally similar \cite{ishikawa2021mapping,2023ApJ...946...38M}.

\paragraph*{Complexity of magnetic structure} The correlation between the large-scale magnetic fields, as shown in Table 1 (more details in Table S2), can characterize the complexity of magnetic structure in the upper layers of the star during the marked epoch, or even considering the decade-long observations (see SI 1B and Figure S6-S9). The difference between the large-scale fields is larger, as indicated by correlation coefficients approaching zero. YZ CMi displays distinct sinusoidal rotational modulation and exhibits the strongest correlation between photospheric and chromospheric fields among our observations, suggesting a more organized and stable magnetic field topology. The modulation patterns and correlations among different layers in the other two stars, EV Lac and AD Leo, exhibit more complex characteristics. In EV Lac, $\BcL$ also shows a substantial correlation with $\Bp$, similar to YZ CMi. However, there is a notable negative correlation between $\BcM$ and the fields in the lower layers. AD Leo exhibits weaker correlations between the layers; magnetic fields in both chromospheric layers appear to be anti-correlated with the photospheric field during the marked epoch, while $\BcL$ and $\BcM$ are highly correlated. These observations suggest that YZ CMi has a more organized large-scale magnetic field structure than the other two stars (SI 1C). Previous theoretical investigations showed that more complex magnetic fields can support less prominence mass on M-dwarfs, resulting in weaker stellar winds \cite{2019MNRAS.482.2853J,2021MNRAS.505.5104W}. AD Leo and YZ CMi are both believed to be fully convective stars with a strong dipole magnetic field \cite{2023A&A...676A..56B,shulyak2017strong}. Although the topologies are similar, YZ CMi has $>30$ times stronger stellar wind than AD Leo\cite{2021ApJ...915...37W}. The difference in stellar winds from these M-dwarfs appears to be consistent with our evaluation of magnetic field complexity.

\paragraph*{Relationship to flare occurrence and energetics} The occurrence frequency of solar flares has been found to be related to the complexity of magnetic field structures in active regions \citep{2019LRSP...16....3T}. From observations of the Transiting Exoplanet Survey Satellite (TESS), we found that the flare frequency distributions on these stars likely relate to the complexity of the magnetic field (SI 1D and Figure S10). The correlation analysis described above suggests that AD Leo has the most complex large-scale magnetic field among the three stars. We notice that, compared to the other two stars, AD Leo has fewer low-energy flares and more high-energy flares (especially superflares). The difference in stellar winds from these M-dwarfs appears to be consistent with our evaluation of magnetic field complexity. This suggests that the occurrence frequency of superflares might be related to the complexity of large-scale magnetic fields on M-dwarfs. The potential explanation is that a richer small-scale field on the surface may produce a more complex field, which can lead to increased magnetic energy dissipation (thus more large flares and fewer small flares) and prune the flare frequency distribution.

\section*{Discussion}
Our observations reveal stable large-scale magnetic fields in the lower and middle chromospheres of active M-dwarfs at a month-long timescale. The longitudinal magnetic fields at different atmospheric layers are generally correlated, which aligns well with those in solar observations\cite{2023ApJ...946...38M, ishikawa2021mapping,2013A&ARv..21...66S}. Given the cancellation effect of opposite-polarity fields during disk integration, the average chromospheric fields could be as strong as the photospheric fields. The high magnetic field strengths we measured in the chromosphere, comparable to photospheric values, coupled with the complex structural variations identified across atmospheric layers, suggest that diffuse or even more complex magnetic field structures dominate the chromospheric topology of these active M-dwarfs. This finding aligns with observational evidence that these stars harbor abundant small-scale magnetic structures throughout their atmospheres\cite{2020A&A...635A.142K,2021A&ARv..29....1K}. This scenario also indicates that a substantial amount of magnetic energy is stored in the upper atmospheres of these stars, which may be released through flares, coronal mass ejections, and intense UV and X-ray radiation \cite{linsky2017stellar,hall2008stellar}. Such frequent eruptions and enhanced radiation likely have a large impact on the start and sustainability of life on nearby orbiting exoplanets \cite{2016NatGe...9..452A,linsky2019host}.

These findings provide critical constraints for understanding the energy budget responsible for heating and eruptions in the stellar upper atmospheres, thus enabling evaluation of the impact of various types of stellar magnetic activity on the habitability of surrounding exoplanets. Considering that chromospheric magnetic fields of stars other than the Sun have never been measured before, our study contributes observations that can inform investigations of magnetic fields in stellar upper atmospheres.

\section*{Methods}
\paragraph*{Spectropolarimetric observations}
The Echelle SpectroPolarimetric Device for the Observation of Stars (ESPaDOnS) is an instrument that includes an achromatic polarimeter. This device has been stationed at the Cassegrain focus of the 3.6m Canada-France-Hawaii Telescope at the top of Mauna Kea in Hawaii, USA, since early 2005 \cite{donati2006espadons}. Meanwhile, NARVAL (not an acronym) is a stellar spectropolarimeter that was designed based on ESPaDOnS and was adapted to the specifics of the 2m Télescope Bernard Lyot (TBL), situated atop Pic du Midi in southwest France since 2006\cite{arnaud2003stellar}. Under polarimetric mode, ESPaDOnS and NARVAL provide complete coverage of the 370 to 1050 nm wavelength range and reach a resolution of R = 65,000 in a single exposure. During this run, polarimetric exposures are divided into four consecutive subexposures. Each subexposure collects data at different angles of the two half-wave rotatable Fresnel rhombs in the polarimetric module. This sequence is designed to eliminate spurious polarization signatures at first-order\cite{semel1993zeeman}. In the standard procedures to extract the spectropolarimetric data, “Null” spectra are obtained by pair-processing sub-exposures corresponding to identical azimuths of the retarding plate or Cassegrain bonnette \cite{donati1997spectropolarimetric}, in parallel with polarized spectra. These “Null” spectra are not expected to contain any signal and should be pure noise if there is no spurious polarized signature in the $V$ profile. We checked our $V$ profile if any significant signal can be seen in the “Null” profile and whether it is similar to the $V$ profile in order to exclude spurious signatures. The total efficiency of the instrument, including the spectrograph and CCD detector, is around 12\%. The software package Libre-ESpRIT\cite{donati1997spectropolarimetric,donati2006surprising} is used for extracting the normalized, reduced data for unpolarized and polarized spectra corresponding to each observing sequence. All reduced spectra used in this work can be extracted from the PolarBase database\cite{petit2014polarbase}. 

We only used the data from 2005-2019 observed with polarimetric mode with the peak S/N of the Stokes $V$ signal over 60. The basic stellar parameters of the selected M-dwarfs used in this work are taken from the study dedicated to their large-scale surface magnetic topologies\cite{morin2008large}. The observation time is phased with the ephemeris: $\mathrm{HJD} = \mathrm{HJD0} + P_\mathrm{rot}E$, where the zero point of heliocentric Julian dates is chosen equal to HJD0=2453950.0 and the rotational periods have been estimated from Zeeman-Doppler Imaging (ZDI) inversion\cite{morin2008large}. Due to the Earth's orbit, ground-based observations for our targets have only a few-month window per year, known as an epoch. A detailed observation log, including the information, can be found in Source Data. 

\paragraph*{Magnetic fields measurement from photosphere to middle chromosphere}
Using the Least Squares Deconvolution (LSD, detailed below) method \cite{donati1997spectropolarimetric}, we extracted average photospheric information from thousands of photospheric spectral lines and produced mean Stokes $I$ and $V$ profiles for all the unpolarized and polarized spectra, respectively. The longitudinal (line-of-sight projected) component of the stellar-disk integrated photospheric magnetic field $\Bp$ was estimated from the first-order moment of the Stokes $V$ profile based on the center-of-gravity (COG) method\cite{1979A&A....74....1R, donati1997spectropolarimetric} decribed below. The cores of the Calcium II infrared triplet (Ca IRT) at wavelengths 849.8 nm, 854.2 nm, and 866.2 nm, typically formed in the middle chromosphere\cite{linsky2017stellar,2023ApJ...954...73H}, were analyzed. We applied the LSD method to these lines and subtracted the photospheric components to obtain the pure chromospheric Stokes $I$ and $V$ profiles (Figure S1). The mean longitudinal magnetic field in the middle chromosphere $\BcM$ was estimated from these profiles using the COG method. Simultaneously, we measured the lower chromospheric longitudinal magnetic field $\BcL$ from the pure emitting hydrogen H$\alpha$ line at 656.3 nm (details in SI 1A \& Figure S2)\cite{2019A&A...623A.136H,2023ApJ...946...38M}. 

\paragraph*{Mean line profiles derived by the Least-Squares Deconvolution method}
Least-squares deconvolution (LSD) is a cross-correlation method, enabling the acquisition of average unpolarized and polarized line profiles with an improved signal-to-noise ratio. This enhancement uses spectral features formed in approximately the same disc region and height \cite{donati1997spectropolarimetric}. Line lists for photospheric analysis are generated from spectrum synthesis via model atmospheres extracted from the Vienna Atomic Line Database (VALD) \cite{piskunov1995vald}. We predominantly employed stronger lines corresponding to the photospheric absorption of an early-to-mid M spectral type star, aligning with the characteristics of our targets. Lines that deviate from the average behavior of the photosphere, particularly chromospheric emissions and telluric lines, are excluded from the list. Approximately 5,000 intermediate to strong atomic absorption spectral lines are simultaneously utilized to extract the average polarization information in line profiles. This approach yields typical noise levels of about 0.06\% (relative to the unpolarized continuum level) per 1.8 km/s velocity bin, resulting in a multiplex gain in Stokes $V$ S/N of $\sim10$ for the photospheric mask. We applied the same approach to the three lines of Ca IRT  and obtained an S/N gain of $\sim1.6$. LSD is not applied to the H$\alpha$ line. We analyze H$\alpha$ individually since its broad Balmer wings break the LSD narrow-line approximation.

\paragraph*{Mean longitudinal magnetic field}
In observations of these low-mass stars, we obtain only \emph{disk-integrated}
Stokes profiles. Consequently, the quantity we derive with the center-of-gravity
(COG) technique is the longitudinal field, $\Bl$, i.e., the first moment of the
circular polarization (Stokes $V$) profile that represents the line-of-sight (LOS) integral
over the visible stellar hemisphere. Because it is an LOS projection,
$\Bl$ can take positive or negative values depending on whether the net signed
flux points toward or away from the observer; its magnitude should not be
interpreted as the absolute field strength at a specific atmospheric height.
COG method is applicable for both the spatially resolved and unresolved profiles and $\Bl$ can be obtained from the first moment of the normalized Stokes $V$\cite{1979A&A....74....1R, donati1997spectropolarimetric,2004......deglinnocenti}:
\begin{equation}
    \Bl = -2.14 \times 10^{-11} \frac{\int \upsilon V(\upsilon) d\upsilon}{\lambda_0 g_\mathrm{eff} c \int [I_c-I(\upsilon)]d\upsilon}
\end{equation}
,where $\nu$ is the velocity to the line center, $\lambda_0$ is the wavelength of the spectral line, $c$ is the speed of light and $g_\mathrm{eff}$ is the effective Land\'e factor. Note that this formula works for a magnetic field (kilo-Gauss) that is even stronger than that in the case of the weak field approximation (WFA), which assumes that the Zeeman splitting is significantly smaller than the line's Doppler width. Under WFA, circular polarization signatures are proportional to $\Bl$ in the first order and can have the same estimation of $\Bl$ as the COG method for weak fields with a similar expression\cite{1979A&A....74....1R}. In a strong field, WFA breaks. However, the COG method is still valid.  
The corresponding uncertainty mainly results from the photon noise of Stokes $I$ and $V$ profiles. This formula is used to estimate the mean photospheric longitudinal field from the mean photospheric line profiles obtained using the LSD method, with typical $g_\mathrm{eff}=1.24$ and $\lambda_0=650$nm.
In the chromosphere, the COG method is still valid and can be applied to chromospheric spectral lines\cite{2023A&A...678A.163A}. In our work, we take $g_\mathrm{eff} = 1$ and $\lambda_0=656.28$nm for the H$\alpha$ line. For the LSD profiles of Ca IRT formed in the middle chromosphere, we have typical values of $g_\mathrm{eff} = 0.968$ and $\lambda_0=858.56$nm, taking from the LSD process, and the $g_\mathrm{eff}$ of each line is taken from the VALD.

\paragraph*{Subtract the photospheric components from CaIRT line profiles}
Information about the middle chromosphere is extracted from the Ca IRT.  Both the Stokes $I$ and $V$ signals of these three lines can be contaminated by photospheric signals from the stellar disk \cite{2007MNRAS.380.1297D}. Considering the weak signal and similar profile shapes of the triplet, we applied the LSD method to the Ca IRT and obtained the mean Ca IRT Stokes $I$ and $V$ profiles. Assuming the wing of the broad absorption band (slightly outside the emission core) anchors to the photosphere in unpolarized Stokes $I$ profiles, we use a Lorentzian model to fit the photospheric component\cite{2007MNRAS.380.1297D}. This component has a broad enough Doppler profile, where the Zeeman splitting is negligible, and WFA is applicable (see Methods: Mean longitudinal magnetic field). Under the first-order, the corresponding photospheric model in circular polarized Stokes $V$ profile is a derivation of the Lorentzian model multiplied by a fixed coefficient of $-4.67\times10^{-12}g_\mathrm{eff}\Bp \lambda_0  c$, where $\Bp$ is the mean longitudinal photospheric field estimated from the LSD. Figure S1 shows an example for modeling the photospheric components in Stokes $I$\&$V$. The residual profiles for both Stokes $I$\&$V$ are used to estimate the mean longitudinal middle chromospheric field $\BcM$. The emission core of Ca IRT should be purely from the chromosphere, and we measured its magnetic fields with the COG method.

\section*{Data Availability}
Data from Figure 2 and the observation journal are provided in Source Data. The original spectral data can be accessed through the PolarBase database \cite{petit2014polarbase} at \url{https://www.polarbase.ovgso.fr}.

\section*{Code Availability}
The codes used to generate Figures 1 and 2 are available on CodeOcean \cite{Cang2025code}. Additionally, a Python package for performing Least Squares Deconvolution (LSD) on spectra is accessible at \url{https://github.com/folsomcp/LSDpy}, as a part of SpecpolFlow\cite{Folsom2025}.

\clearpage
\section*{Acknowledgments}
T.C. acknowledges funding from the support of the National Natural Science Foundation of China (NSFC) through grant 12503035. T.C. and W.Z. acknowledge funding from the support of the NSFC through grant 12273002. H.T. acknowledges the support of NSFC (grants 12425301/12250006) and the New Cornerstone Science Foundation through the Xplorer Prize. J-F.D. acknowledges funding from the European Research Council (ERC) under the H2020 research \& innovation program (grant agreement 740651 NewWorlds). JNF acknowledges the support from the National Natural Science
Foundation of China (NSFC) through the grants 12090040/12090042
/12427804 and supported by the China Manned Space Program with grant
no. CMS-CSST-2025-A013, and the Central Guidance for Local Science
and Technology Development Fund under No. ZYYD2025QY27. H.L. acknowledges the support of the National Key R\&D Program of China (2021YFA1600500/ 2021YFA1600503). HP.L. acknowledges funding from the support of NSFC through grant 12103004. S.B. acknowledges funding from the Dutch Research Council (NWO) under the project "Exo-space weather and contemporaneous signatures of star-planet interactions" (with project number OCENW.M.22.215 of the research programme "Open Competition Domain Science- M"). T.C is supported by the LAMOST fellowship as a Youth Researcher, which is funded by the Special Funding for Advanced Users, budgeted and administered by the Center for Astronomical Mega-Science, Chinese Academy of Sciences (CAMS), and the China Postdoctoral Science Foundation (2023M730297).

\section*{Author Contribution}
TQ.C analyzed the spectral data, identified the polarized chromospheric signal in M-dwarfs, and wrote the manuscript. P.P., J-F.D., H.T., and TQ.C. designed the methods of magnetic field analysis. TQ.C, P.P., J-F.D., A.L.A., and JN.F contributed to the initial interpretation of spectral lines. P.P., J-F.D., J.M., and S.B. contributed to the original collection of spectra and the analysis of the photospheric magnetic field and contributed to the manuscript writing. H.T. substantially contributed to the analysis and interpretation of the chromospheric magnetic field and assisted in writing the manuscript. H.L. and HP.L. contributed to the theoretical model of the chromosphere as a confirmation of the feasibility of the magnetic field analysis method and contributed to the manuscript writing. XY.M., XY.H, KY.X., and WK.Z. contributed to the correlation analysis and illustration of the magnetic fields.

\section*{Competing Interests}
The authors declare no competing interests.

\clearpage
\section*{Tables}
\begin{table*}[h]
    \centering
        \caption{\textbf{Correlation coefficients ($\rho$) of magnetic fields between layers (P for photosphere, cL and cM for lower and middle chromosphere).} The coefficients are derived from all data, and a marked epoch with full phase coverage for each star.} 
    \begin{tabular}{c|cc|cc|cc}
       Star  &  \multicolumn{2}{c|}{$\rho_\mathrm{(P,cL)}$}  &\multicolumn{2}{c|}{$\rho_\mathrm{(P,cM)}$}  &\multicolumn{2}{c}{$\rho_\mathrm{(cL,cM)}$}  \\
       & Marked & All & Marked & All & Marked & All \\
           \hline
      AD Leo   &  -0.22 & +0.26 & -0.57 & -0.19 & +0.83 & +0.64 \\
      \hline
      YZ CMi & +0.84  & +0.76 & +0.65 & +0.36 & +0.79 & +0.76 \\
      \hline
      EV Lac  & +0.84 & +0.69 & -0.45 & -0.35 & -0.57 & -0.18  \\
      \hline
    \end{tabular}
    \label{tab:corr_main}
\end{table*}

\clearpage
\section*{Figures}
\begin{figure*}[ht]
    \centering
    \includegraphics[width=1\textwidth]{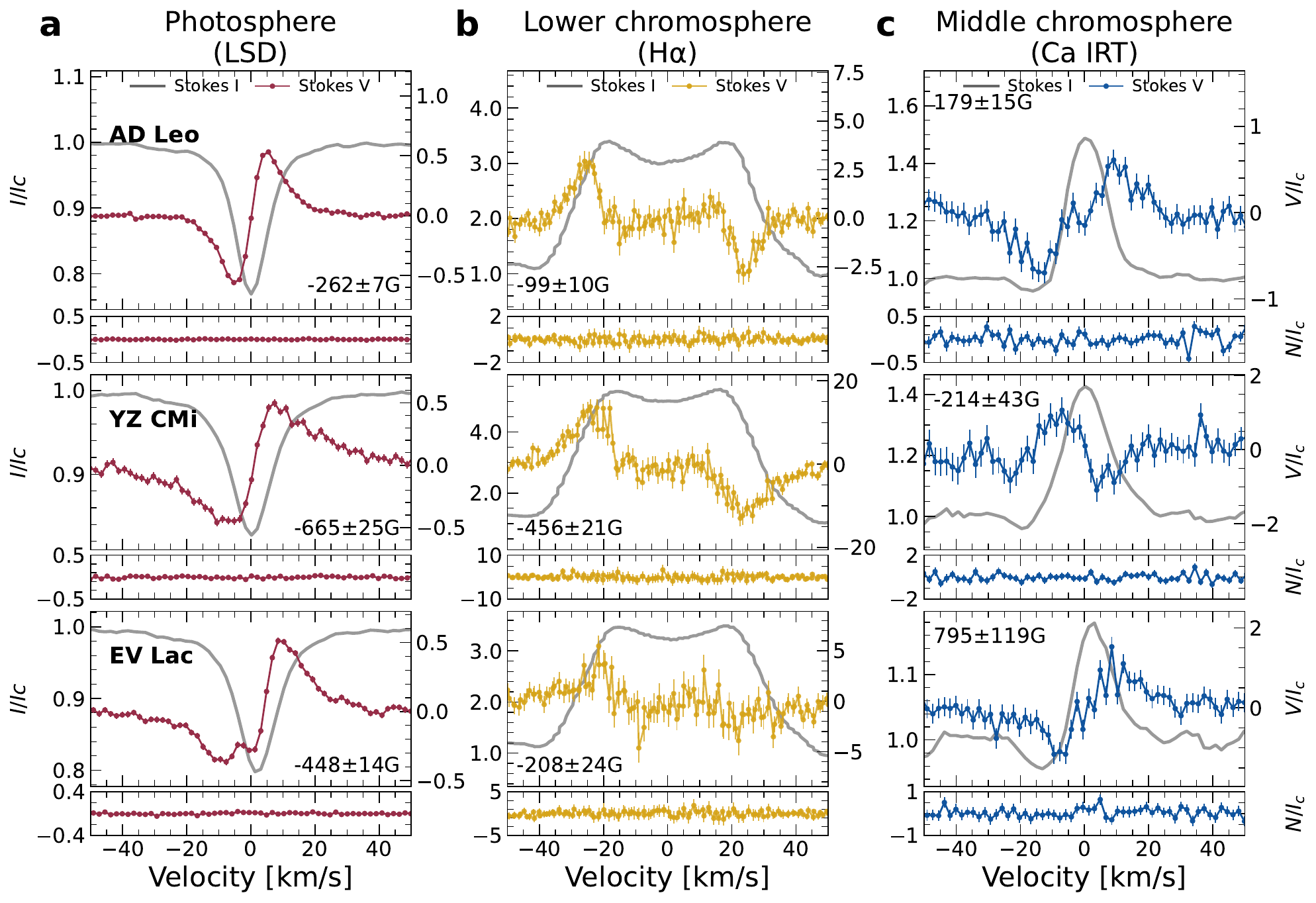}
    \caption{\textbf{Examples of spectropolarimetric spectral lines of three stars.} The profiles are sampled from the observations of 2007Jun25 (AD Leo), 2007Jan27(YZ CMi), 2005Sep18 (EV Lac), with above-average signal-to-noise. \textbf{a.} The spectral profiles of LSD line profile (photosphere). \textbf{b.} H$\alpha$ (lower chromosphere). \textbf{c.} Ca IRT (middle chromosphere). The gray lines show the unpolarized intensity profiles normalized to the unpolarized continuum intensity $I_c$. The color lines represent Stokes $V$ profiles normalized to $I_c$, and the corresponding Null ($N/I_c$) profile with no expected signal from the targets. Both $V/I_c$ and $N/I_c$ are multiplied by 100 for the purpose of illustration. The error bars of $V/I_c$ indicate the $\pm1\sigma$ uncertainties. The velocities have been corrected from the stellar system radial velocity to shift the line center of photospheric LSD to 0 km/s. The inferred magnetic field is printed in each panel. }
    \label{fig:line}
\end{figure*}

\begin{figure*}[ht]
    \centering
    \includegraphics[width=1\textwidth]{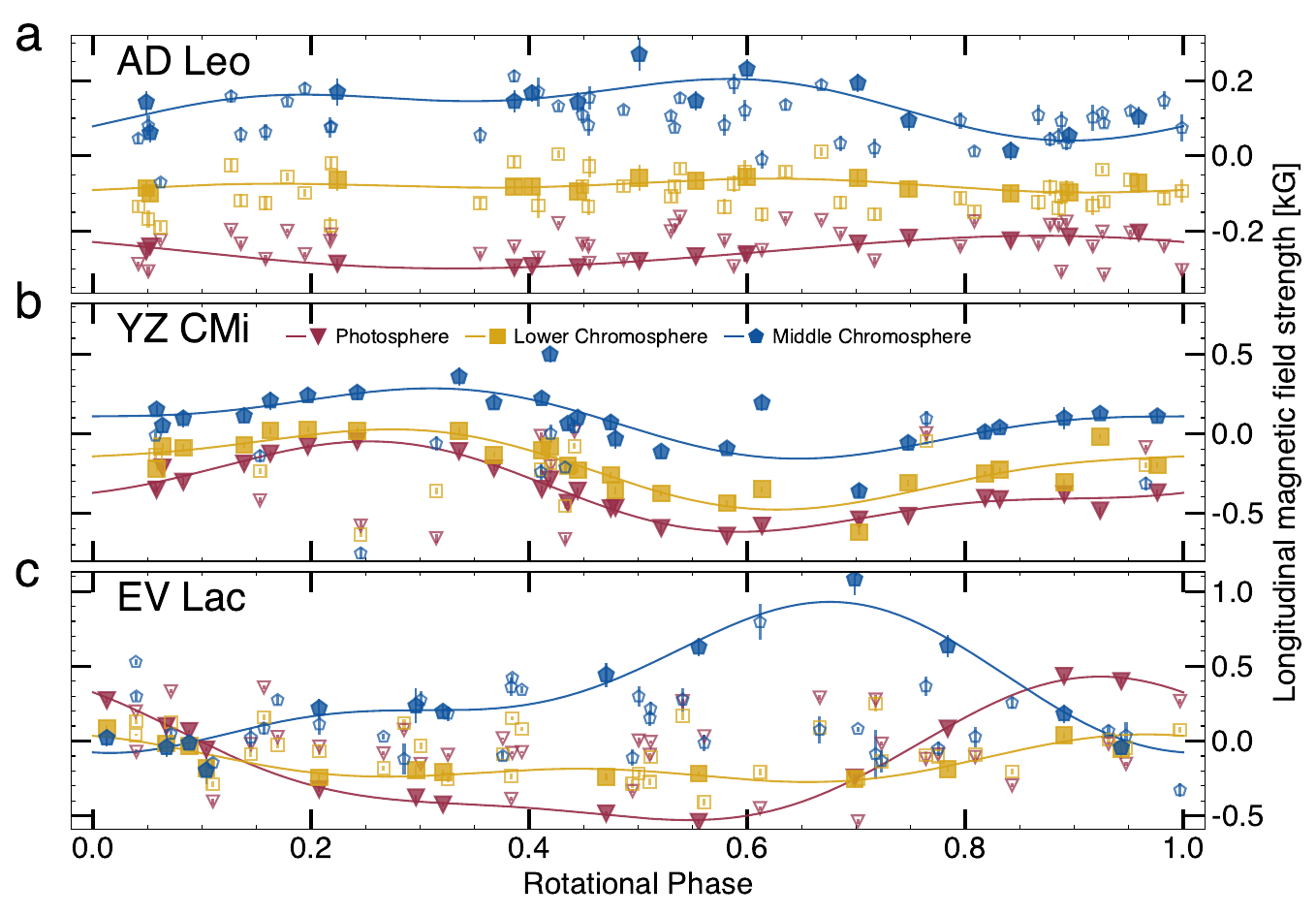}
    \caption{\textbf{Phase-folded mean longitudinal magnetic field strengths in the photospheres and chromospheres.} \textbf{a.}AD Leo. \textbf{b.}YZ CMi. \textbf{c.}EV Lac. \textbf{a-c.}The rotational phases were derived from the same ephemera. Red triangles are photospheric fields $\Bp$ inferred from LSD profiles; yellow squares are lower chromospheric fields $\BcL$ from H$\alpha$, blue pentagons are middle chromospheric fields $\BcM$ from Ca IRT. Filled markers are taken from a marked epoch with full phase coverage for each star. For these epochs, the linear Pearson correlation coefficients between the field strengths of two layers are also listed in the left block. For each marked epoch, we give a two-harmonic Fourier series to fit the pattern, showing the periodic signal of the magnetic field variation.}
    \label{fig:Bl}
\end{figure*}

\section*{Supplementary Information}
\paragraph*{1. Supplementary Text}
\paragraph*{1A. Formation heights of H$\alpha$ and Ca IRT lines in the chromosphere}
All three targets in this study have small $v\sin i$ values (Table S1). Consequently, the unpolarized, disk-integrated profiles we observed closely approximate the mean profiles derived from all disk pixels. We observed that all H$\alpha$ profiles display a well-known two-peak structure, consistent with predictions from Non-local Thermodynamic Equilibrium (NLTE) atmospheric models \cite{2005A&A...439.1137F, 2023ApJ...954...73H}.

For solar active regions, longitudinal magnetic fields detected through H$\alpha$ line wings originate from the bottom of the chromosphere, whereas the line core emanates from the mid-to-upper chromosphere, similar to the core of Ca IRT \cite{2023ApJ...946...38M}. Figure S2 shows an example of the two-peak structure of H$\alpha$ in Stokes I and corresponding V profiles. We use standard indicators for the structure: the line-center depression (H$\alpha$3), the violet and red emission peaks (H$\alpha$2v and H$\alpha$2r), and the flux minima (H$\alpha$1v and H$\alpha$1r). The wings of the profile are between H$\alpha$1x and H$\alpha$2x (x can be v or r). 

To evaluate the strength of the circular polarization signal, we calculated the average absolute ratio between $V$ and $I$ profiles ($|V/I|$) for both the wings and core of the line profile. We found that the average $|V/I|$ of the wings is at least twice that of the core, suggesting that the Stokes $V$ signal in H$\alpha$ is predominantly detected within the wings of the Stokes $I$ profiles. Hence, the detected magnetic field in the H$\alpha$ profiles should be attributed to a lower height in the chromosphere.

According to solar observations \cite{2023ApJ...946...38M}, the core of Ca IRT is formed in the middle chromosphere and its polarization profiles can be used to detect the fields there.

\paragraph*{1B. Variation of the measured longitudinal magnetic field with height}
The variation patterns of the longitudinal magnetic field $\Bl$ measurements shown in Figure 2 are predominantly influenced by the large-scale magnetic field structures\cite{morin2008large}. As shown in Figure S3, the measurements of YZ CMi in the selected epoch generally form one repeating pattern in the phase-folded diagram, indicating that the large-scale magnetic fields are generally stable in one epoch by neglecting the differential rotation of the stars. For AD Leo EV Lac, their patterns are less clear than YZ CMi, but they can be fitted by a double sine function as shown in Figure 2. The large-scale magnetic fields in the photosphere of all three targets generally exhibited a dipole structure\cite{morin2008large, 2023A&A...676A..56B, 2024arXiv240308590B}. However, the magnetic axis for these stars was not precisely aligned with the rotational axis, particularly for YZ CMi and EV Lac. Additionally, the inclination angle between the rotational axis and the line-of-sight from Earth is significantly larger for YZ CMi and EV Lac compared to AD Leo (Table S1). 

Figure S4 presents possible scenarios that may explain the field with opposite polarities at different heights. We see two magnetic field lines originating from the same region but reaching different locations. For one large-scale magnetic loop, if the direction of the loop sharply changes with height, the measured longitudinal field at different heights can have different polarities. These scenarios may explain the different strengths of average longitudinal magnetic fields at different heights, and the sign of the middle chromospheric field can be opposite to that of the photospheric field. As a test, we applied the photospheric field of AD Leo in Potential Field Source Surface (PFSS) modeling\cite{2020JOSS....5.2732S} to extrapolate the large-scale field at a typical height of the chromosphere of $1.003R_*$ according to the Sun. As shown in Figure S5, with the inclination of $i=20^\circ$ and $60^\circ$, although the radial field has a very tiny variation, it is still possible to result in a decrease of $\Bl$ or even an opposite sign. We also made simple simulations (details in Supplement Text 1C) to reproduce the observed anti-correlations between layers. Note that the chromosphere should be much thinner than shown in the figures, meaning that large-scale magnetic loops are highly curved and lie very close to the stellar surfaces. Although actual situations are much more complex, these toy models can already explain our observed field variation with height, highlighting the strongly varying magnetic topologies with height.

Figures S6 to S8 present correlation analyses of all our measurements of longitudinal fields between different atmospheric layers, along with corresponding linear fittings for each correlation. Although the correlations between longitudinal fields in different layers vary within each epoch, there is an overall global correlation with all observations. For instance, in panel (a) of Figure S8, the photospheric magnetic field strengths of EV Lac in 2016 are weak but aligned with the linear fitting derived from the 2007 measurements, which exhibits a broader range of field variations. Similarly, the single observation of YZ CMi in 2009 fits the trend observed in 2007. Even though the field strength of AD Leo is relatively stable, a universal slope can still be observed in Figure S6 (c). These linear relationships suggest an intrinsic connection between the large-scale magnetic field structures from layer to layer, e.g., sharing the same magnetic loops. The numerical Pearson coefficient $r$ and its two-tailed $p$-value are listed in Table S5, indicating a highly significant correlation for all data between photospheric and chromospheric fields, especially for YZ CMi and EV Lac.

\paragraph*{1C. Complexity of the large-scale magnetic field structure}
The complexity of the large-scale magnetic field close to the surface in our work is evaluated by the correlation between layers. Zeeman-Doppler Imaging (ZDI) reconstruction of the 2-D chromospheric magnetic field map can potentially provide information on field complexity. It is unclear whether the small number of available chromospheric lines can be used to create a reliable map. Nevertheless, some ideas from ZDI can still be useful for testing the field complexity. For a model of magnetic field topology, we can define the complexity of the large-scale magnetic field by the degree ($\ell$) of the spherical harmonics decomposition of the field \cite{2006MNRAS.370..629D}. The radial field is defined by $B_r\ (\theta,\phi)=\Sigma_{\ell=1}^L\Sigma_{m=0}^\ell \mathrm{Re}[\alpha _{\ell m} Y_{\ell m}(\theta,\phi)]$, where $(\theta,\phi)$ are the colatitude and longitude on the stellar surface, $Y_{\ell m}=c_{\ell m}P_{\ell m}(\mathrm{cos}\theta)e^{im\phi}$ with $c_{\ell m}=\sqrt{(2\ell+1(\ell-m)!)/(4\pi(\ell+m)!)}$ and  $P_{\ell m}$ the associated Legendre polynomial. $m$ gives the order of the spherical harmonics mode and describes the azimuthal variation or symmetry of the field. In practice, $\alpha_{\ell m}$ is a complex coefficient corresponding to the radial poloidal field with specific $\ell$ and $m$. We examined dipole ($\ell=1$), quadrupole ($\ell=2$), and octupole ($\ell=3$) fields, constructed from a photospheric field with slight asymmetry (give equal energy to $m$=0 and $m>0$) and extrapolated to the upper layers via Potential Field Source Surface (PFSS) modeling\cite{2020JOSS....5.2732S}. We set an inclination of $i=40^\circ$ (Figure S9), the correlation between the B$_\ell$ of the photosphere and a layer outside at 1.05 $R_*$ is high with $\ell=1$ and 2, but moderate for $\ell$=3. These extrapolations indicate that a more complex magnetic field structure results in a lower correlation between magnetic fields at different layers. Our complexity ranking for the targets is as follows: AD Leo is the most complex, YZ CMi is the most organized, and EV Lac is intermediate.  
These simulations can also be used to explain the opposite polarities and anti-correlation between different layers in our observations. In Figure S9 (b), opposite polarity frequently happens, indicating that the quadrupole of the magnetic field is enough to generate this phenomenon. In Figure S9 (c), we show a possible situation with an anti-correlated pattern together with opposite polarities between the surface and outer layers, suggesting that the octupole mode can account for both two phenomena.
Note that the solar chromosphere is located at only around 0.3\% R$_\odot$ above the surface, which is about an order of magnitude lower than the layer we take in the simulation. This is because the topology of the extrapolated field with PFSS does not vary too much from the surface to such a low height, resulting in a correlation coefficient value close to 1. The magnetic loops in the model (black lines in the middle panels of Figure S9) need to be at a higher layer to achieve the degree of bending to reproduce the phenomena.

\paragraph*{1D. Flare Frequency Distributions}
Flare Frequency Distributions (FFDs) describe the relationship between the occurring frequency of stellar flares and their energy or intensity. Specifically, FFDs quantify how often flares of varying energies occur over a given period. We extracted the flares from TESS Sector 1-57, with the state-of-the-art method dedicated to TESS flare detection developed by \cite{2024ApJS..271...57X}. Figure S9 presents FFDs for our three targets, using data from the Transiting Exoplanet Survey Satellite (TESS) from 2018 to 2024, including three sectors for each star. We see a trend that stars with more low-energy flares have fewer high-energy flares. Figure S9 shows that stars with more complex fields tend to exhibit more high-energy flares and fewer low-energy ones. In the solar case, it is well-known that more complex magnetic field structures in active regions produce more flares\cite{2021ApJ...915...37W}, which is similar to the occurrence of high-energy flares in our observations. Fewer low-energy flares in the situation of more complex fields can be attributed to the rich small-scale field on the surface of these M dwarfs, which can convert the low-energy flares to increased magnetic dissipation and prune the FFD.

\newpage
\paragraph*{2. Supplementary Figures}

\paragraph*{S1. Modeling the photospheric component of Ca IRT in Stokes $I$ \& $V$} (a) A Lorentzian model (blue curve) is used to fit the broad wing (20-100 km/s) of the Ca IRT LSD Stokes $I$ profile (black dots). (c) The residual I profile (subtraction of the Lorentzian model from the Stokes I profile) is used for $\BcM$ estimation. (b) The corresponding circular polarization model (blue curve) with fixed $\Bp$. The black dots with error bars are the measured Stokes V profile. (d) The residual V profile (subtraction of the circular polarization model from the Stokes V profile) is used for $\BcM$ estimation.

\begin{figure}[h]
    \centering
    \includegraphics[width=1\textwidth]{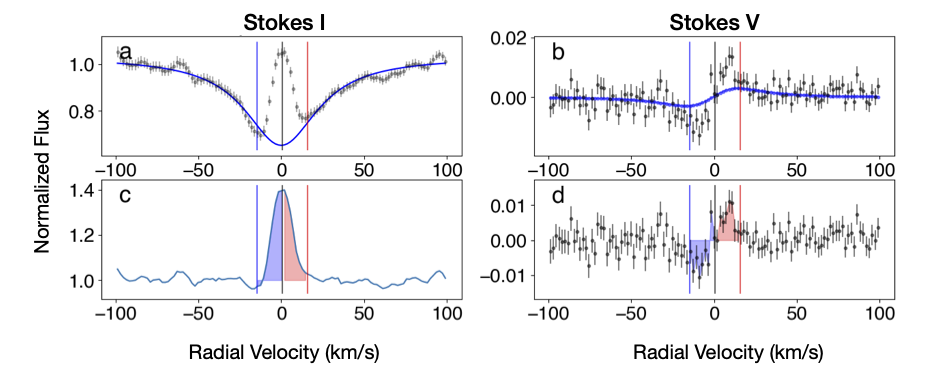}
\end{figure}

\paragraph*{S2.An example of the Stokes-I and V profiles of YZ CMi from the observation on 2007-12-29.} Blue and red vertical lines indicate the positions of emission peaks H$\alpha$2v and H$\alpha$2r. 

\begin{figure}[h]
    \centering
    \includegraphics[width=0.8\textwidth]{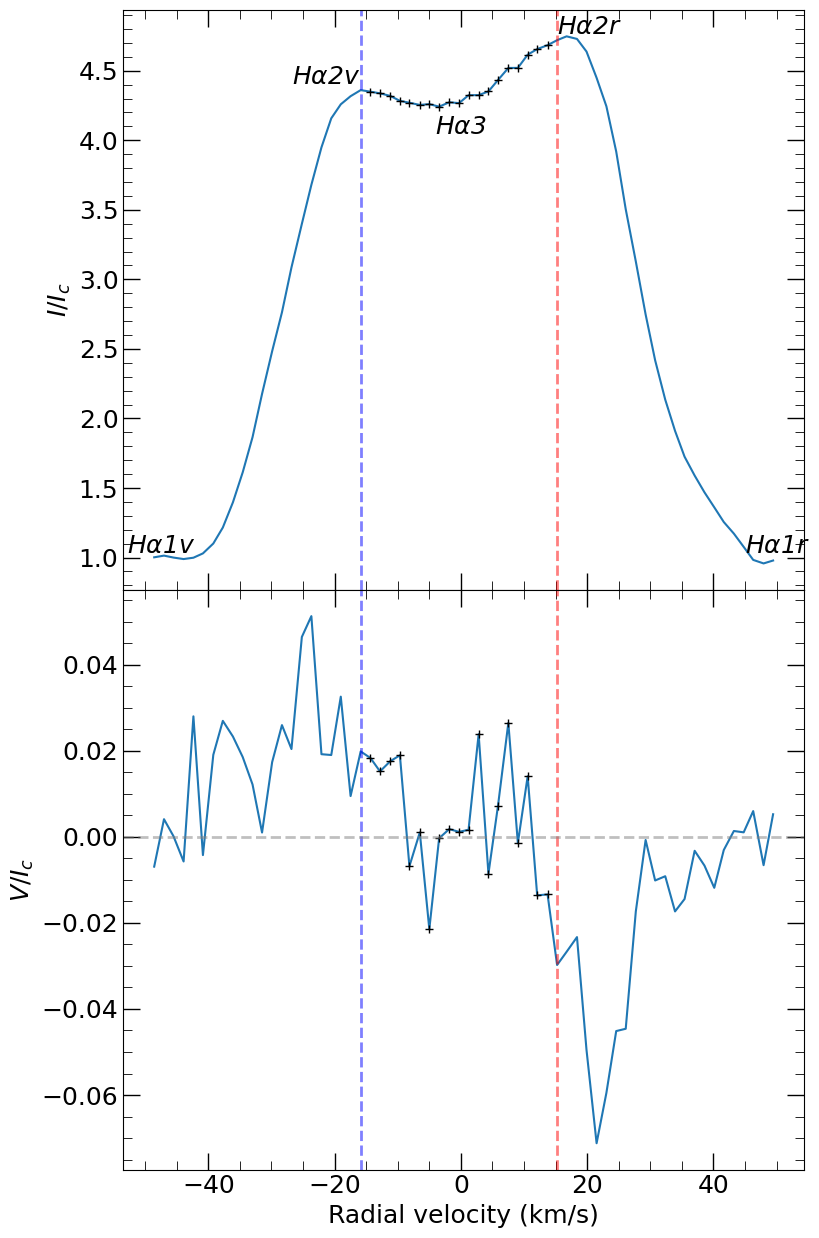}
\end{figure}

\clearpage

\clearpage
\paragraph*{S3. Repeating pattern of measurements in the selected epoch of YZ CMi.} The measurements are the same as in Fig. 2; the color shows the day difference from HJD = 2454500. For reference, the solid line represents the fit of the data using a double sine function. 

\begin{figure}[h]
    \centering
    \includegraphics[width=1\textwidth]{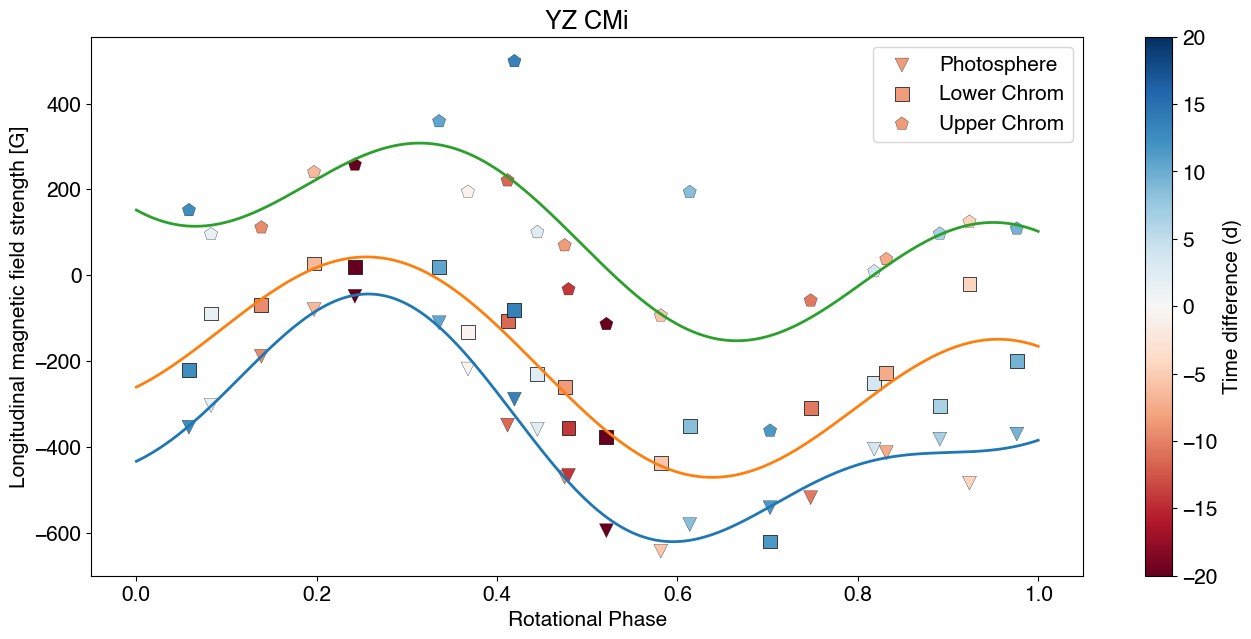}
\end{figure}

\clearpage
\paragraph*{S4. A cartoon showing a possible scenario of magnetic field with opposite polarities at different heights}
For example, we take a minimized magnetic field structure on the surface of AD Leo. Two magnetic field lines originate from region A and then reach B and C, respectively. The layers with red, orange and blue colors indicate the photosphere, lower chromosphere and middle chromosphere, respectively. The angle between the line-of-sight and rotation axis is set to be $i=20^{\circ}$, with the shadow part of the star invisible for observation. The visible parts of magnetic field lines are shown in the same colors as layers. The arrows with different colors indicate the polarities of the measured magnetic fields at corresponding layers.
\begin{figure}[h]
    \centering
    \includegraphics[width=1\textwidth]{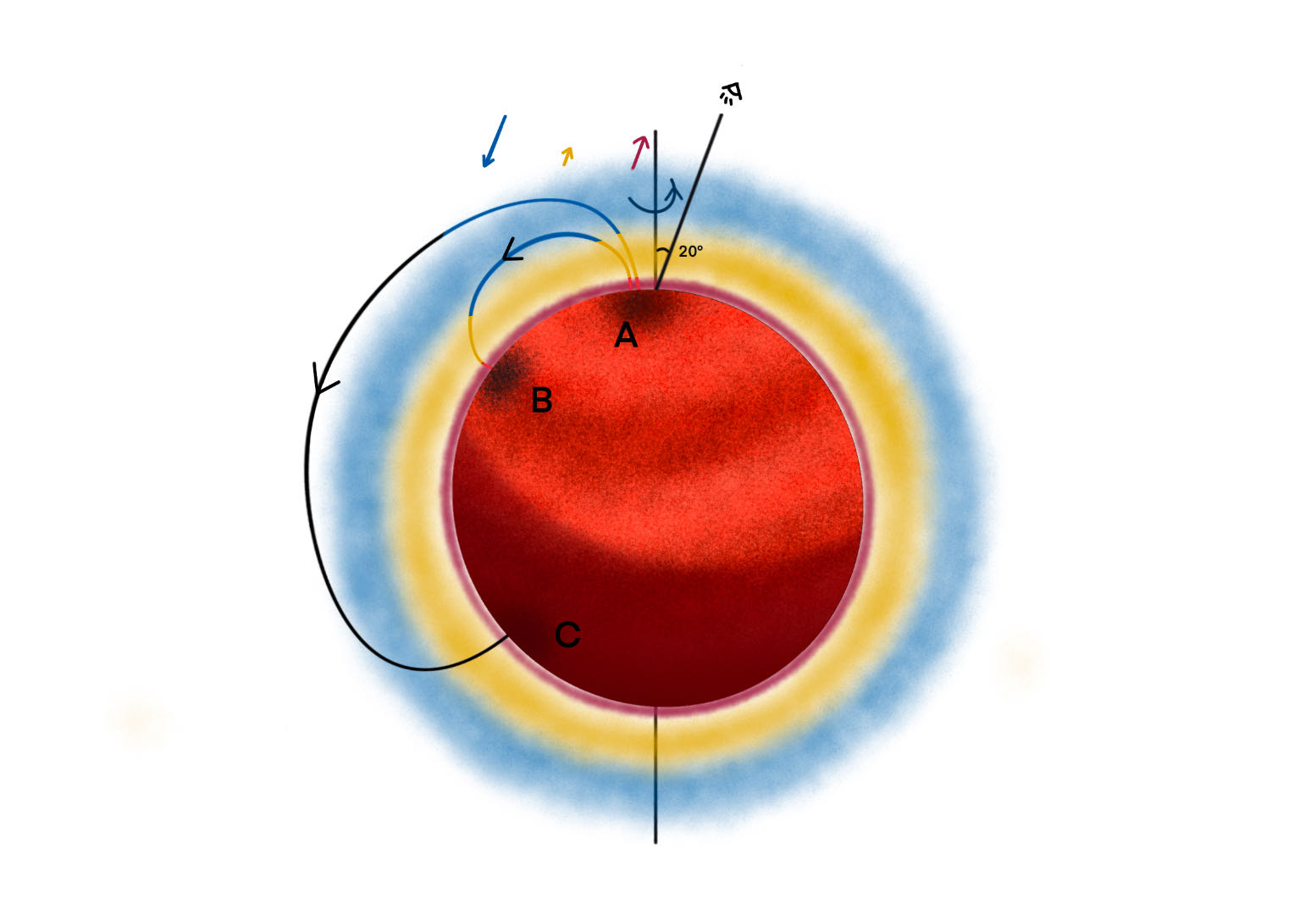}
\end{figure}

\clearpage
\paragraph*{S5. PFSS modeling experiment for AD Leo}
We take the radial field of AD Leo in (a) for PFSS modeling, and obtain the extrapolated field at $1.003R_*$ in (b). The corresponding line-of-sight mean field $\Bl$ vs. rotation phase with different inclinations of $i=20^{\circ}$ and $60^{\circ}$ are shown in (c) and (d), respectively. 
\begin{figure}[h]
    \centering
    \includegraphics[width=1\textwidth]{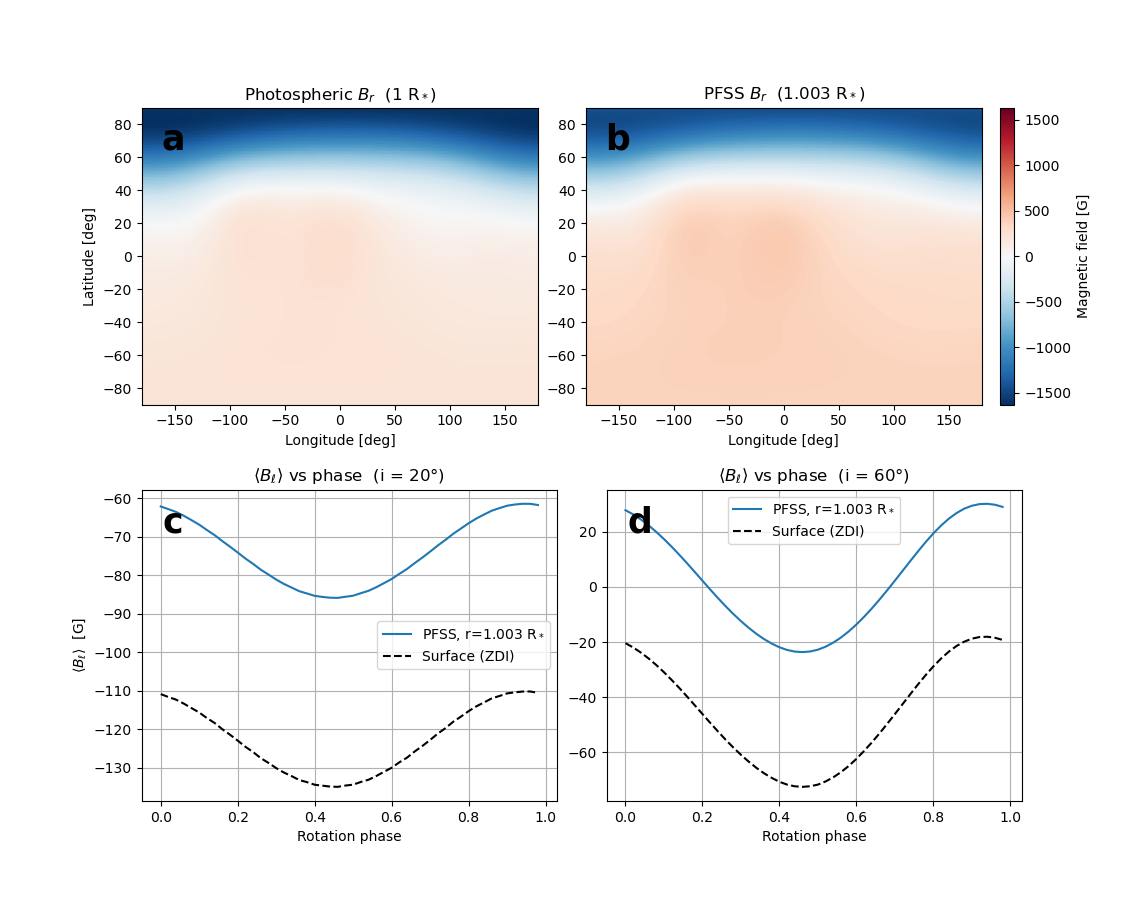}
\end{figure}

\clearpage
\paragraph*{S6. Correlation among photospheric, lower and middle chromospheric longitudinal magnetic fields for AD Leo}
(a) The correlation between $\Bp$ and $\BcL$ for all data. The original values and corresponding uncertainties can be found in Table S4. Each epoch legend has the start date and is shown with a different color. For the epochs with over three observations, a linear regression is presented by a line with the same color. A linear regression for all observations is presented with the gray dashed line. (b,c) Same as panel (a), but for the correlation between $\BcM$ and $\Bp$ (b), $\BcL$ (c), respectively. The corresponding Pearson correlation coefficient for each linear fitting is presented in the table in the upper-right panel with the same color selection as the other panel for each epoch.

\begin{figure}[h]
    \centering
    \includegraphics[width=1\textwidth]{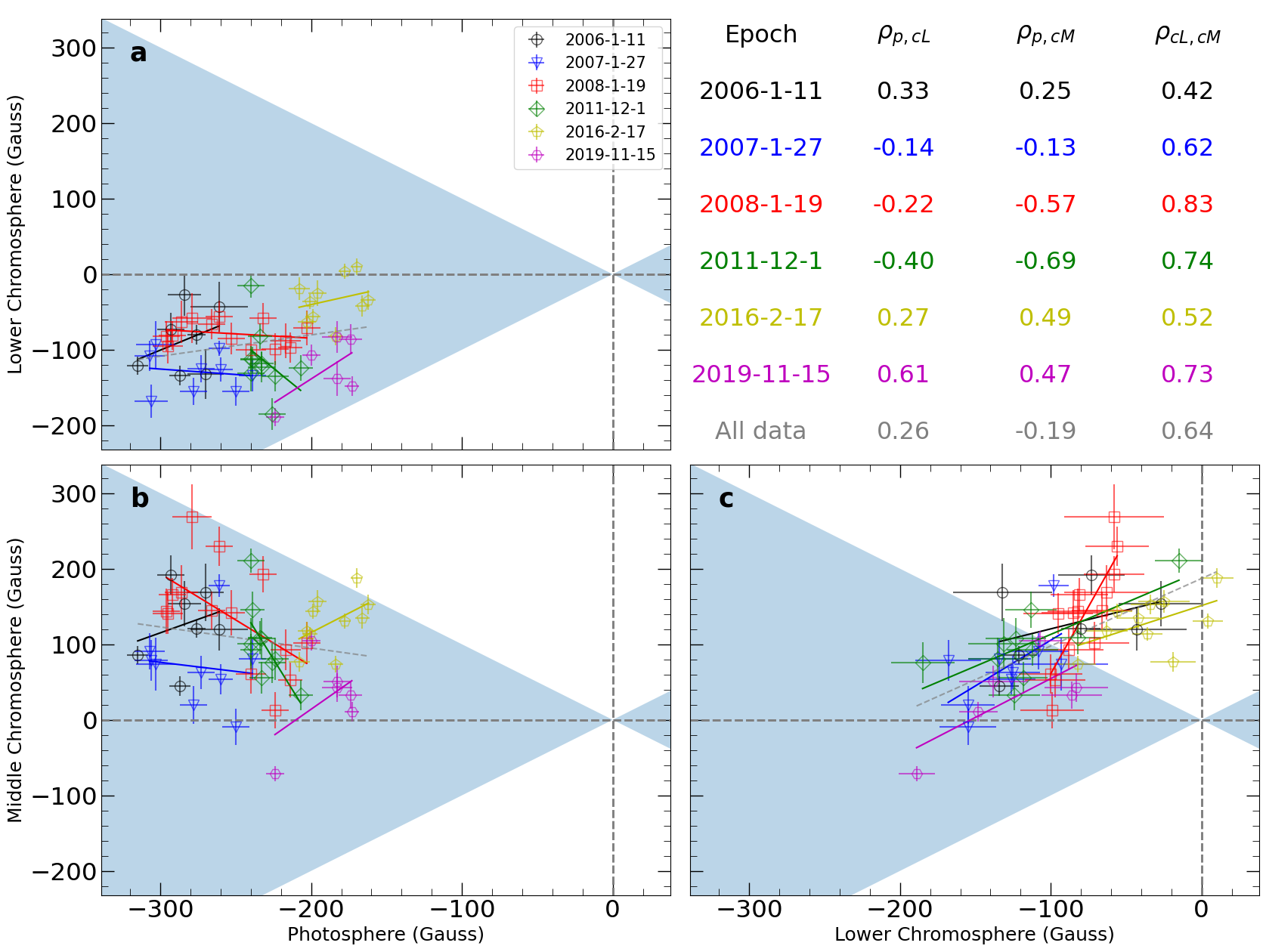}
\end{figure}

\clearpage
\paragraph*{S7. Same as Fig. S6 but for YZ CMi.}
The original values and corresponding uncertainties can be found in Table S2.
\begin{figure}[h]
    \centering
    \includegraphics[width=1\textwidth]{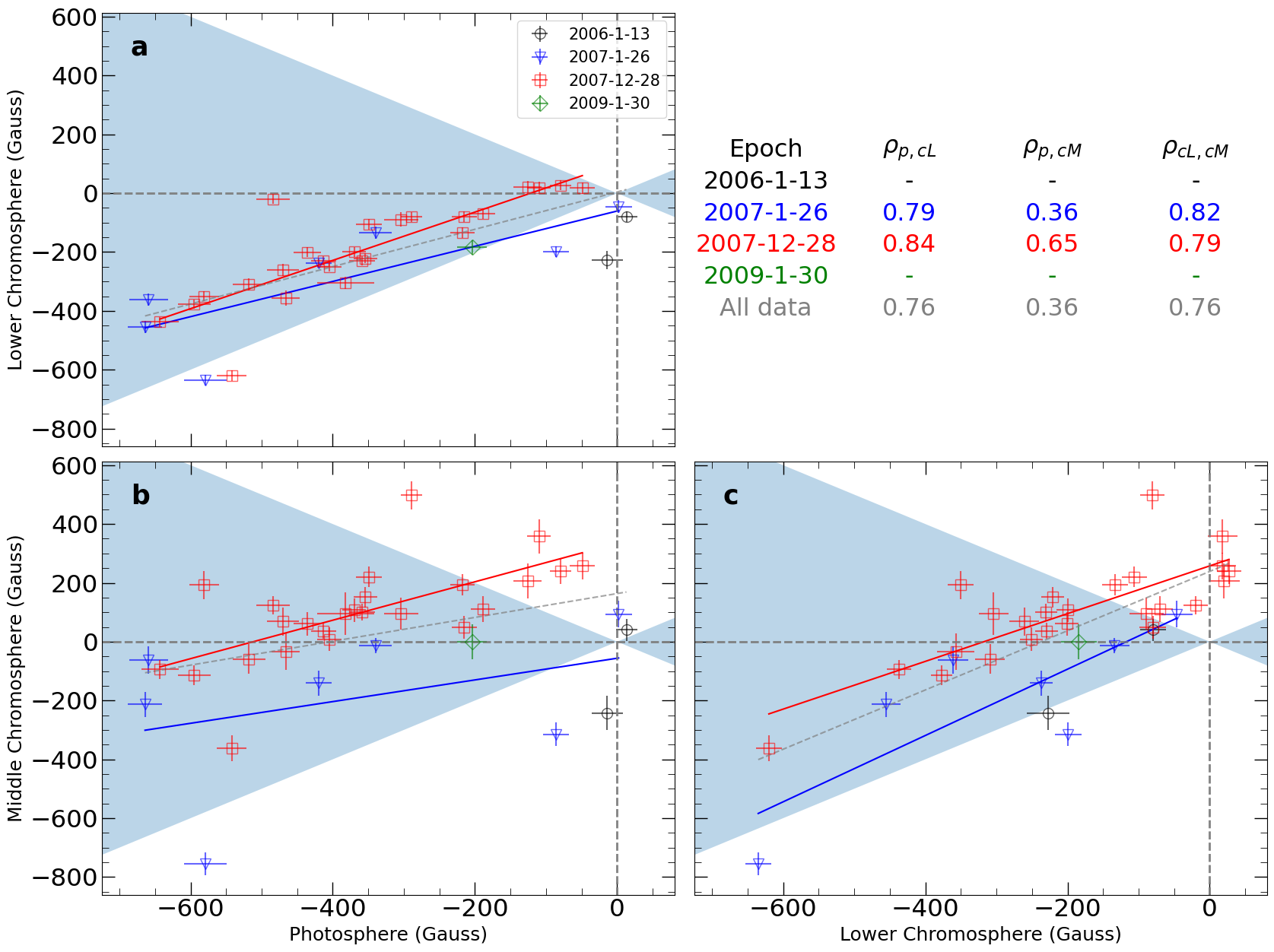}
\end{figure}

\clearpage
\paragraph*{S8. Same as Fig. S6, but for EV Lac.}
The original values and corresponding uncertainties can be found in Table S3.
\begin{figure}[h]
    \centering
    \includegraphics[width=1\textwidth]{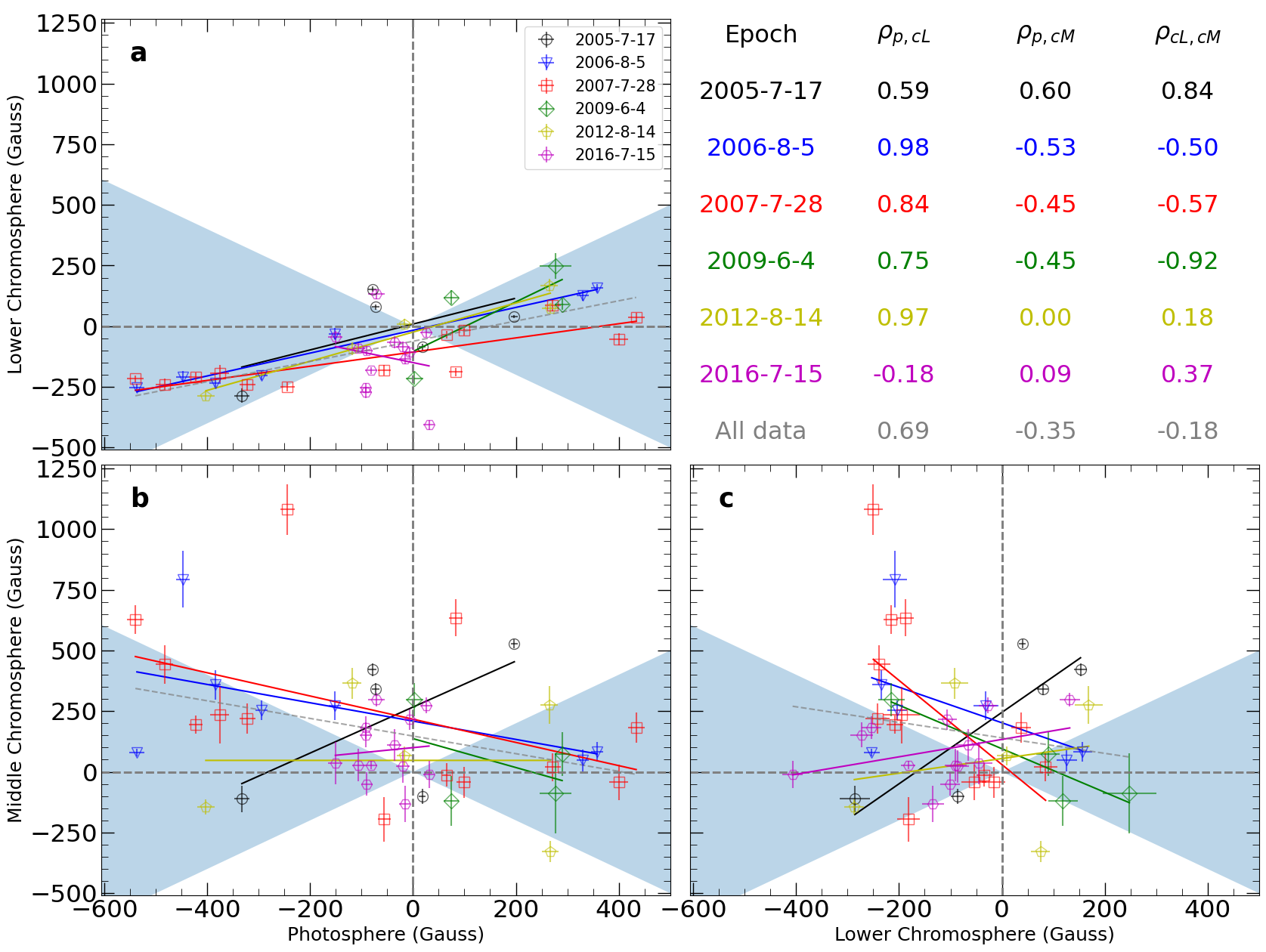}
\end{figure}

\clearpage
\paragraph*{S9. PFSS modeling with $\ell =1$, 2, and 3}
Panels on the left show radial field distribution on the photosphere and corresponding PFSS modeling results. The blue and red areas indicate the strengths of the magnetic field, with the color bar shown in Gauss. Blue and red lines represent the negative- and positive-polarity open field lines, respectively. The black lines are closed field lines. Panels on the right present the variation of the average line-of-sight magnetic field strength of the photosphere (black dashed lines, $R=1R_*$) and a higher layer (blue lines, $R=1.05R_*$) as a function of the rotation phase. The corresponding correlation coefficients $\rho$ between the magnetic fields of the two layers are presented on the left panels.
\begin{figure}[h]
    \centering
    \includegraphics[trim={0 0 5cm 0},clip,width=1\textwidth]{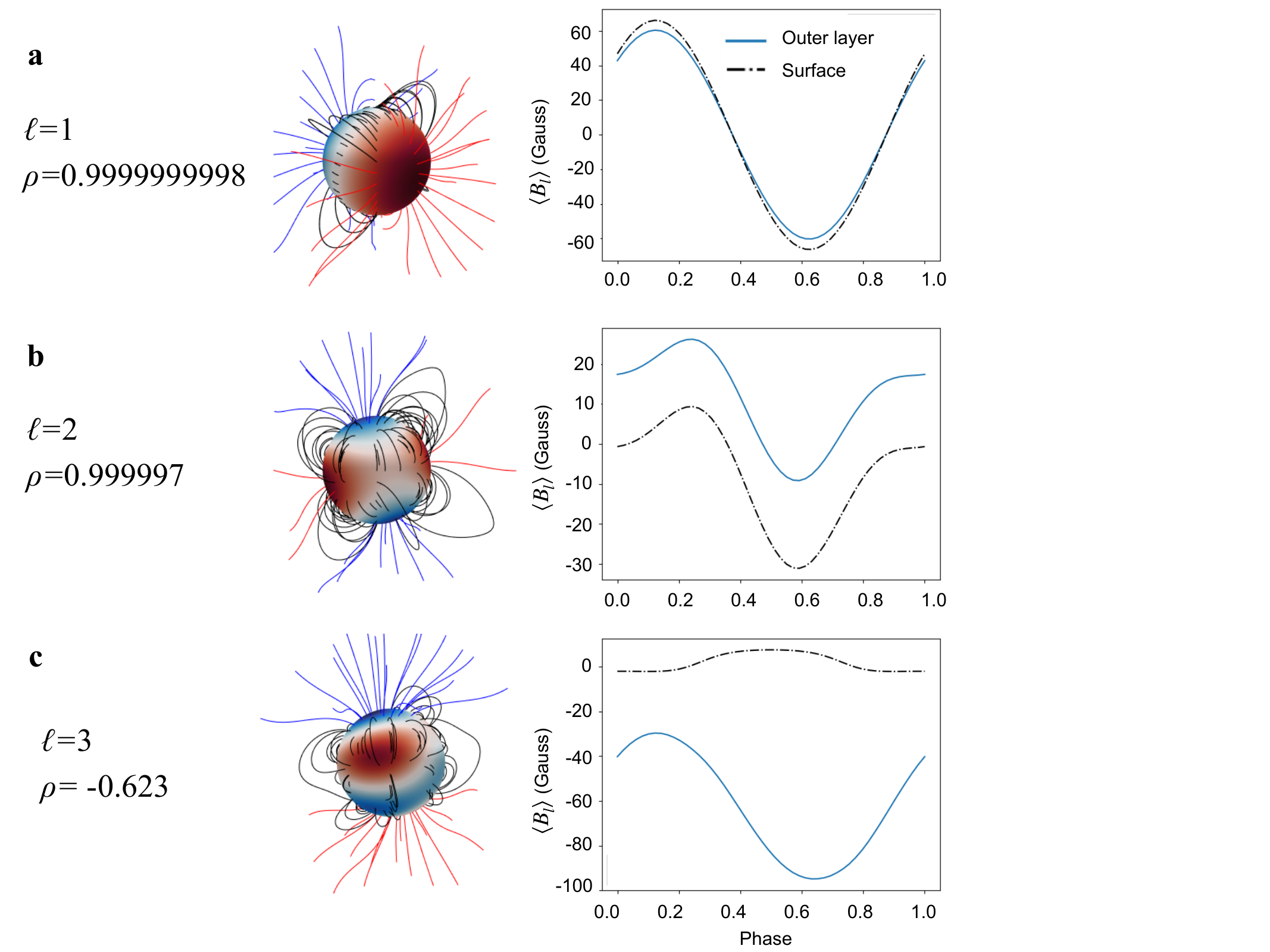}
\end{figure}

\clearpage
\paragraph*{S10. Cumulative Flare Frequency Distributions of the three stars, derived from TESS observations.}
The corresponding slope (red lines) $\beta$ gives the power-law fitting for the middle-to-high energy part ($10^{32}-10^{33.5}$ erg), the value can be found in Table S1.
\begin{figure}[h]
    \centering
    \includegraphics[width=1\textwidth]{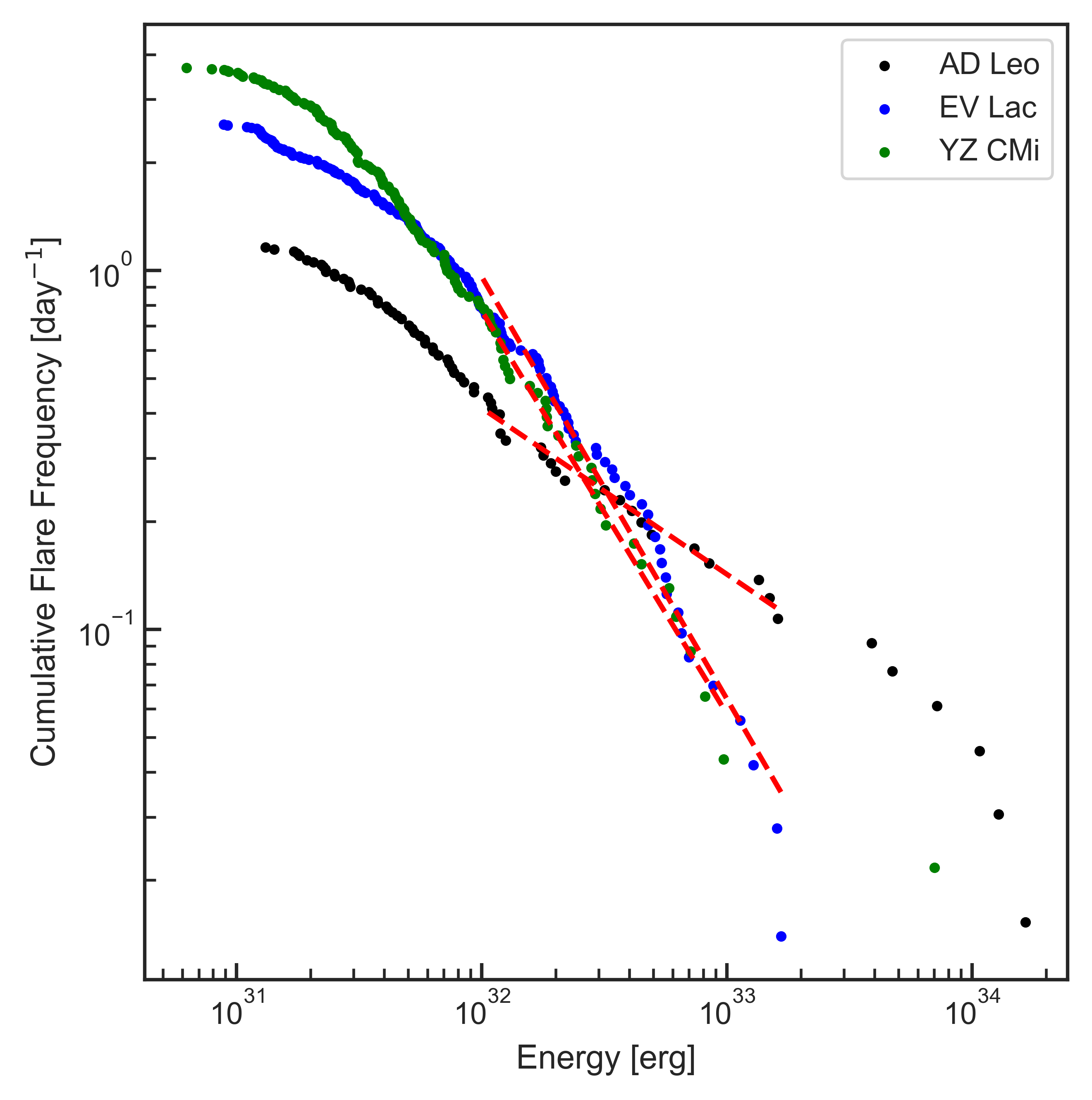}
\end{figure}

\clearpage
\paragraph*{3. Supplementary Tables}
\paragraph*{S1. Stellar parameters of the observed M-dwarfs} Most of the parameters are taken from \cite{morin2008large}, including stellar mass, radius, spectral type, period, radial velocity (RV), $v\mathrm{sin}~i$, and the inclination angle (INCL) between the rotational axis and the line-of-sight from Earth. Stellar age are taken from \cite{2024ApJ...960...62E}, the percentage of large-scale field $\langle B_V \rangle / \langle B_I \rangle $ are taken from \cite{2020A&A...635A.142K}. $\beta$ gives the slope of power-law fitting of FFD (Fig. S10).

\begin{table*}[h]
    \centering
    \begin{tabular}{lrrr}
    \hline
       Values & AD Leo & YZ CMi & EV Lac  \\
         \hline
Mass ($M_\odot$) &0.42 &0.31 &0.32 \\
Radius ($R_\odot$) & 0.38 &0.29 & 0.30\\
Spec Type &M3& M4.5 & M3.5 \\
log(Age) [yr]  &-0.764 & -0.092 & -0.052 \\
Period (d) &2.22 & 2.77 & 4.378 \\
RV (km/s)& 12.3 & 26.6 & -1.5 \\
$v\mathrm{sin}~i$ (km s$^{-1}$) & 3.0 & 5.2 &4.0 \\
INCL ($^\circ$)&20 & 60 & 60 \\
$\langle B_V \rangle / \langle B_I \rangle $&  5.8\%  & 13.3\% & 10.2\% \\ 
$\beta$& $-0.79\pm0.09$ & $-1.024\pm0.0016$ & $-0.67\pm0.09$ \\
\hline
    \end{tabular}
    \label{tab:info}
\end{table*}

\clearpage
\paragraph*{S2. Epoch-by-epoch statistical information of each star}
For each observing epoch defined in Tables S2-S4, the correlations between the disk-integrated longitudinal magnetic fields measured in the three atmospheric layers. Column 1 gives the UTC start date of the epoch and Column 2 its end date; Column 3 lists the number of individual Stokes $V$ spectra acquired during that epoch (ObsCount).  Columns 4–6 provide the Pearson correlation coefficient \(r_{\rm p}\) between the epoch-averaged fields of two layers—photosphere (ph), lower chromosphere (cL) and middle chromosphere (cM)—with the two-tailed probability \(p\). Columns 7–9 list the epoch-averaged longitudinal magnetic fields themselves, in gauss, for the photosphere (mean $\Bp$), the lower chromosphere (mean $\BcL$) and the middle chromosphere (mean $\BcM$), respectively.  Within each stellar block an “Overall” line combines all epochs of that star and gives the correlation coefficients and mean fields derived from the full decade-long data set.

\begin{table*}[h]
    \centering
\resizebox{0.9\textwidth}{!}{
\begin{tabular}{llc|rrr|rrr}
\toprule
StartDate & EndDate & ObsCount & rp(ph,cL) & rp(ph,cM) & rp(cL,cM) & mean($\Bp$) & mean($\BcL$) & mean($\BcM$) \\
\midrule
AD Leo\\
2006-1-11 & 2006-6-12 & 7 & (0.33,0.46) & (0.25,0.59) & (0.42,0.35) & -284 & -87 & 127 \\
2007-1-27 & 2007-6-25 & 9 & (-0.14,0.71) & (-0.13,0.74) & (0.62,0.08) & -275 & -129 & 70 \\
2008-1-19 & 2008-2-16 & 14 & (-0.22,0.45) & (-0.57,0.03) & (0.83,0.00) & -254 & -78 & 137 \\
2011-12-1 & 2012-1-16 & 10 & (-0.40,0.26) & (-0.69,0.03) & (0.74,0.02) & -231 & -114 & 101 \\
2016-2-17 & 2016-4-20 & 10 & (0.27,0.46) & (0.49,0.15) & (0.52,0.12) & -187 & -34 & 129 \\
2019-11-15 & 2019-11-21 & 6 & (0.61,0.20) & (0.47,0.34) & (0.73,0.10) & -190 & -125 & 29 \\
Overall & Overall & 56 & (0.26,0.05) & (-0.19,0.17) & (0.64,0.00) & -238 & -91 & 106 \\
\midrule
YZ CMi\\
2006-1-13 & 2006-2-7 & 2 & - & - & - & 0 & -153 & -100 \\
2007-1-26 & 2007-2-8 & 7 & (0.79,0.03) & (0.36,0.43) & (0.82,0.02) & -392 & -295 & -201 \\
2007-12-28 & 2008-2-16 & 25 & (0.84,0.00) & (0.65,0.00) & (0.79,0.00) & -358 & -193 & 100 \\
2009-1-30 & 2009-1-30 & 1 & - & - & - & -204 & -184 & -1 \\
Overall & Overall & 35 & (0.76,0.00) & (0.36,0.03) & (0.76,0.00) & -340 & -211 & 26 \\
\midrule
EV Lac\\
2005-7-17 & 2005-9-18 & 5 & (0.59,0.29) & (0.60,0.28) & (0.84,0.07) & -54 & -19 & 216 \\
2006-8-5 & 2006-8-12 & 7 & (0.98,0.00) & (-0.53,0.22) & (-0.50,0.26) & -161 & -92 & 270 \\
2007-7-28 & 2007-8-18 & 13 & (0.84,0.00) & (-0.45,0.12) & (-0.57,0.04) & -83 & -131 & 257 \\
2009-6-4 & 2009-8-15 & 4 & (0.75,0.25) & (-0.45,0.55) & (-0.92,0.08) & 161 & 60 & 41 \\
2012-8-14 & 2012-8-22 & 5 & (0.97,0.01) & (0.00,1.00) & (0.18,0.77) & -1 & -24 & 47 \\
2016-7-15 & 2016-8-12 & 13 & (-0.18,0.55) & (0.09,0.77) & (0.37,0.21) & -54 & -126 & 88 \\
Overall & Overall & 47 & (0.69,0.00) & (-0.35,0.02) & (-0.18,0.22) & -54 & -84 & 167 \\
\midrule
\end{tabular}
}
\end{table*}

\bibliography{sn-bibliography.bib}

\end{document}